\documentstyle[epsfig]{elsart}
\begin{document}
\begin{frontmatter}
\title{Halo Dark Matter and Ultra-High Energy Cosmic Rays}
\author{Pasquale Blasi and Ravi K. Sheth}
\address{NASA/Fermilab Astrophysics Group, Fermi National Accelerator
Laboratory, Box 500, Batavia, IL 60510-0500, USA}
\maketitle

\begin{abstract}
The decay of very heavy metastable relics of the Early Universe can 
produce ultra-high energy cosmic rays (UHECRs) in the halo of our own 
Galaxy. 
On distance scales of the order of the halo size, energy losses are 
negligible---no Greisen-Zatsepin-Kuzmin cutoff is expected. 
In this letter we show that, as a consequence of the hierarchical 
build up of the halo, this scenario predicts the existence of
small scale anisotropies in the arrival directions of UHECRs. 
We also suggest some consequences of this scenario which will
be testable with upcoming experiments, as Auger.
\end{abstract}
\end{frontmatter}

Heavy particles ($m_X\sim 10^{12}-10^{14}$ GeV) can be produced in the 
Early Universe in different ways
\cite{kolb,bere,kuzmin,kt} and their lifetime can be finite though very
long compared to the present age of the universe. The decay of these
particles results in the production of UHECRs, as widely discussed in the
literature \cite{bere,sarkar,bbv,blasi}.  If the relics cluster in 
galactic halos, as is expected, this can explain the cosmic ray 
observations above $\sim 5\times 10^{19}$ eV.  These particles may 
represent an appreciable fraction of the cold dark matter in the universe 
\cite{kolb,bere,kuzmin,kt}.
Can this class of models explain the details of the present data,
including the spectrum, composition and anisotropy of the arrival 
directions? We will briefly review the status of the first two points, 
already discussed in the relevant literature and we will
instead concentrate on the last point, paying particular attention to 
anisotropies one expects on small scales.

The decay of heavy relics results usually in the production of a 
quark-antiquark pair which rapidly hadronizes, generating two jets 
with approximately $95\%$ of the energy in pions, and $\sim 5\%$ in 
baryons. The decay of the pions results in the observed high energy 
particles, mainly in the form of gamma rays, and
in the generation of ultra-high energy neutrinos. The spectrum of the gamma
photons is relatively flat ($\sim E^{-1.5}$) reflecting the behaviour of the
fragmentation function for the quarks. Therefore two main signatures of this
model are: {\it i)} a flat energy spectrum; {\it ii)} composition dominated 
by gamma rays rather than by protons. Moreover, as in all top-down models, 
heavy elements are expected to be completely absent.
Unfortunately present data on the composition is extremely poor and it is
impossible to rule out or confirm the presence of gamma photons in the UHECR
events. 

In \cite{dt,bbv} the issue of the anisotropy was first addressed. 
The anisotropy results from the asymmetric position of the Earth in 
the Galaxy, so that the flux of UHECRs coming from the direction of 
the galactic center should be appreciably larger than
the flux from the anticenter direction. In 
\cite{bm,medina} this issue was considered
more quantitatively, taking into account the exposure of the present 
experiments.
All authors concur that the present data is consistent with 
the predictions of the relic model for practically all reasonable values  
of the model parameters.

Recently an interesting pattern has arisen from the analysis of the 
events with  energy larger than $4\times10^{19}$ eV: in \cite{agasa} 
the sample with this energy cut comprises 47 events, whose overall 
distribution in space does not show appreciable deviation from isotropy.  
However, 3 doublets and one triplet were identified within an 
angular scale of $2.5^o$, comparable with the
angular resolution of the experiment. A complete analysis, 
including the whole set of UHECR events above $4\times 10^{19}$ eV
from the existing experiments was performed in \cite{watson}. 
This extended sample comprises
92 events and shows 12 doublets and two triplets (each triplet is also 
counted as three doublets) within an angle of $3^o$.  
The chance probability of having more than this number of doublets 
was estimated to be $\sim 1.5\%$. Although it is probably too soon to 
rule out the possibility that these multiplets are just a random 
fluctuation, it is instructive to think about the possibility that 
their presence contains some physical information about the sources 
of UHECRs. Clearly the most straightforward possibility
is that the multiplets correspond to some local overdensity in the 
spatial distribution of the sources. Most of the top-down models for 
UHECRs (e.g. strings, necklaces, vortons, etc.) cannot naturally 
explain the multiplets. 

In the following we will discuss how the multiplets can be interpreted in
the context of the super-heavy dark matter (SHDM) model.

The existence of dark matter in galactic halos is all but established,
and the main points are well understood in terms of hierarchical
structure formation scenarios. High resolution N-body simulations, e.g.,   
\cite{simul}, suggest the following: {\it i)} the density 
of dark matter particles in galaxy size halos is peaked in the center; 
the density cusp scales as $\sim r^{-\gamma}$ with $\gamma\sim 1-1.5$ 
on distances $r$ which are much smaller than a core radius, which is of 
the order of several kpc in size; 
{\it ii)} outside the core, the slope of the profile steepens;
it scales as $\propto r^{-3}$ at large distances; 
{\it iii)} the profile is not completely smooth; some of the dark matter  
is in small clumps.
To model the first two of these findings we adopt a dark matter 
density distribution in the form suggested by numerical simulations 
\cite{nfw}:
\begin{equation}
n_H(r)=n^0\frac{(r/r_c)^{-1}}{\left[1+\frac{r}{r_c}\right]^2}
\label{eq:NFW}
\end{equation}
where $r_c$ is the core size and $n^0$ is a normalization parameter. 
These two parameters can be set by requiring that the halo 
contains a given total mass ($M_H$) and that the velocity dispersion 
at some distance from the center is known (in the case of the Galaxy, 
the velocity dispersion is $\sim 200$ km/s in the vicinity of our 
solar system.). 
Alternative fits to the simulated dark matter halos and a discussion of 
whether or not simulated halos appear to be consistent with observations 
are provided in \cite{simul}.

In addition to the smooth dark matter distribution, represented by
eq. (\ref{eq:NFW}), N-body simulations also show that there is a 
clumped component which contains $\sim 10-20\%$ of the total mass. 
The presence of these clumps are a natural consequence of the way in 
which gravity assembles dense virialized halos such as our galaxy 
today from the initially smooth density fluctuation field which 
was present when the cosmic microwave background (CMB) 
decoupled from the baryons.  
Simulations suggest that most of the mass which makes up a 
galactic halo was assembled by merging smaller clumps together 
at about $z\sim 3$.  Much of the mass initially in a small clump which 
falls onto and orbits within the larger halo after that time gets 
tidally stripped from it.  The amount of mass which is lost from 
any given clump increases as the distance of closest approach to 
the galactic halo center decreases; the mass is stripped away, from 
the outside in, as the clump falls towards the center.  
We will call the size of a clump, after its outsides have been stripped 
away, the tidal radius of the clump.  Dynamical friction makes 
the clumps gradually spiral in towards the halo center.  

This has three main consequences.  Firstly, the range of clump masses 
in the halo at the present time is different from that which fell in.  
Secondly, halos of a fixed mass do not all have the same tidal radius.  
Thirdly, the spatial distribution of the clumps in the halo is not the 
same as that of the dark matter.  We found that a good fit to 
the joint distribution in clump mass and position in the simulations 
of \cite{simul} is
\begin{equation}
n_{cl}(r,m) = n_{cl}^0 \left(\frac{m}{M_H}\right)^{-\alpha} 
\left[1+\left(\frac{r}{r_c^{cl}}\right)^2\right]^{-3/2},
\label{eq:clumps}
\end{equation}
where $n_{cl}^0$ is a normalization constant, $r_c^{cl}$ is the 
core of the clumps distribution, and $\alpha$ describes the relative 
numbers of massive to less massive clumps. The simulations suggest that 
$\alpha\sim 1.9$ \cite{simul}.  The constraints on the core size are 
weaker---we will study the range where $r_c^{cl}$ is between 3 and 
30 percent of $R_H$.  
In \cite{simul}, a halo with $M_H\approx 2\times 10^{12}~M_\odot$
contains about $500$ clumps with mass larger than $\sim 10^8~M_\odot$. 
This sets the normalization constant in eq. (\ref{eq:clumps}).

Clumps in the parent NFW halo are truncated at their tidal radii.  
The tidal radius of a clump depends on the clump mass, the density 
profile within the clump, and on how closely to the halo center it 
may have been.  
We assume that clumps of all mass are isothermal spheres (even though 
they are not truly isothermal, \cite{simul} suggest this is reasonably 
accurate):  $\rho_{cl}(r_{cl})\propto 1/r_{cl}^2$, 
where $r_{cl}$ is the radial coordinate measured from the center of 
the clump. The tidal radius of a clump ($R_{cl}$) at a distance $r$ from 
the center of the parent halo is determined by requiring that the 
density in the clump at distance $R_{cl}$ from its center equals the 
local density of the NFW halo at the distance $r$.   
This means that
\begin{equation}
R_{cl} = \left(\frac{m}{4\pi n^0 x_c}\right)^{1/3} x^{1/3} 
\left[ 1+\frac{x}{x_c} \right]^{2/3},
\label{eq:size}
\end{equation}
where $x_c=r_c/R_{H}$, $x=r/R_H$ and $R_H$ is the virial radius of 
the halo, of order $300$ kpc.  The average overdensity within $R_H$ 
is about 200 \cite{simul}.

As shown in \cite{bbv,medina}, the total (energy integrated) flux of 
UHECRs per unit solid angle from a smooth distribution of dark matter 
particles in the halo is:
\begin{equation}
\frac{d\Phi}{d\Omega} \propto \int_0^{R_{max}} dR\, n_H(r(R)),
\label{eq:flux}
\end{equation}
where $R$ is the distance from the detector, and $r$ is the distance 
from the galactic center (so $R$ and $r$ 
are related by trigonometrical relations accounting for the
off-center position of the Earth in the Galaxy). 
The upper limit, $R_{max}$, depends on the line of sight. 

The existence of a clumped component changes the flux in eq. 
(\ref{eq:flux}) only in that $n_H$ should be replaced with the total 
dark matter density, the sum of the smooth and the clumped components. 
It is intuitively obvious that clumped regions will give an excess 
of events from certain directions, as was first pointed out in 
\cite{bereproc}. 

To see how important the clumped contribution is, we used two different 
ways of simulating the observed number of events.  
The first approach consisted of calculating the flux
per unit solid angle [eq. (\ref{eq:flux})] along different lines of 
sight directly, taking into account the smooth plus clumped contributions 
to the total density profile [eqs.~\ref{eq:NFW} and~\ref{eq:clumps}]. 
Once a smooth flux map distribution had been obtained, the UHECR events 
were generated from this distribution. 
In the second approach, the events were generated in two steps.  
First, a random subset of the dark matter distribution, which is supposed 
to represent the subset of particles which decayed, was generated.  
The second step was to draw particles from this distribution, and 
then weight by the probability that the event would actually have been 
detected---so a chosen particle generates an event with probability 
$\propto 1/r^2$, where $r$ is the distance between the particle and 
the detector. 
In both codes, the detector was assumed to be at the position of the
Earth in the Galaxy, and a cut on the directions of arrival was 
introduced to account for the exposure of the AGASA experiment (taken
here as an example).

Fig. 1 shows as an example one of the generated flux maps: 
the map represents
the ratio of the total flux including the contribution from clumps,
to the flux obtained by using a smooth NFW profile.  The various free 
parameters were $r_c=8$ kpc, $r_c^{cl}=10$ kpc, and the mass distribution 
was truncated at a clump mass of $1\%$ of the mass of the NFW halo. 
This sort of plot emphasizes the clump contribution. 
(To avoid confusion many of the smallest scale fluctuations,
which arise from the many low mass clumps, have not been shown.)

To calculate the small scale anisotropies, we generated $10^4$ mock 
samples, each of 92 observed events, and counted the number of 
doublets and triplets for angular scales of 3, 4, and 5 degrees. 
Our codes can also be used to check the corresponding numbers for 
the case of isotropic arrival directions (as in \cite{watson}). Two
sets of values of the cores for the NFW and the clumped component
were adopted, one
in which $r_c=8$ kpc and $r_c^{cl}=10$ kpc (case 1) and the other with 
$r_c=r_c^{cl}=20$ kpc (case 2). The observed numbers of doublets within
3, 4, and 5 degrees for an isotropic distribution of arrival directions
are given in \cite{watson} and are 12, 14 and 20 respectively. 
The number of doublets that we obtain in case 1 are 8, 14, and 21
within 3, 4, and 5 degrees respectively. The probability that 
the number of doublets equals or exceeds that observed is
$12\%$, $47\%$ and $57\%$ respectively.  This should be compared with 
the $1.5\%$, $13.4\%$ and $15.9\%$ quoted in \cite{watson} for an
isotropic distribution of arrival directions. 

We repeated the same calculation for the case 2. The corresponding 
averages and probabilities of exceeding the observed number of 
doublets within 3, 4 and 5 degree scales are 6.6, 12, and 18, 
and $4.5\%$, $29\%$ and $36\%$ respectively. 

In both cases 1 and 2, the number of doublets on angular scales of 
4 and 5 degrees is consistent with the observed values; presumably 
the discrepancy at 3 degrees is random chance. 

We have also studied the occurence of triplets.  
There is some ambiguity as to how a triplet is best defined; we have 
chosen to define triplets as configurations in which all three pairs 
would have been classified as doublets.  (This means, for example, that 
a co-linear configuration of two doublets is not necessarily a triplet.)   
With this definition, the average number of triplets in case one is 
0.5, 1.5 and 3, with the probability of having more than the 
observed triplets (2, 2, 3 respectively) equal to 4\%, 16\% and 35\%.  
For case 2, the correspondent numbers are 0.4, 1, and 2.5 triplets 
and $2\%$, $8\%$ and $20\%$ for the probabilities to have more triplets
than observed.

What is responsible for the multiplet-events in the SHDM model? 
If we study the case in which all the halo mass is in the smooth 
NFW component, then the number of doublets typically drops by 
one or two.  This suggests that the anisotropy due to our position 
in an NFW halo can result in a number of multiplets of events which 
is considerably larger than if the arrivals were from an isotropic 
background.  The number of multiplets from the clumped component is 
mainly affected by the presence of large nearby clumps, whose number 
depends on the high mass cutoff imposed in the mass function of clumps. 
A maximum mass of $1\%$ of the halo mass implies a total mass in the 
clumps of $\sim 10-15\%$ of $M_H$, 
consistent with the results of the simulations \cite{simul}. 
Larger cutoffs imply larger mass fractions, 
which are harder to reconcile with the N-body simulations.

Future experiments will definitely represent the real test for these 
sorts of models. Therefore it is useful to propose tests to be 
performed in the next years, in particular with the upcoming Auger 
experiment \cite{auger}. 
In the following we investigate what a full sky experiment 
would observe if UHECRs were produced by the decay of super-heavy 
halo dark matter.
As discussed in \cite{sommers} a powerful statistical tool that will be
available with a full sky experiment is the angular power spectrum, as a
function of $l$:
\begin{equation}
C(l)=\frac{1}{2l+1} \sum_{m=-l}^{l} a_{lm}^2,
\end{equation}
where the coefficients $a_{lm}$ are defined as
\begin{equation}
a_{lm}=\frac{1}{N_{ev}}\sum_{i=1}^{N_{ev}} Y_{lm}(\theta,\phi),
\end{equation}
and we assumed, for simplicity, that the exposure is flat over all the
sky. Here $N_{ev}$ is the number of events and $Y_{lm}$ are the real valued
spherical harmonics, calculated in the direction determined by the two
angles $\theta$ and $\phi$ of the event.

We simulated 5000 events over all the sky and computed the power spectrum
in two cases: 1) purely NFW profile (no clumps); 2) full dark matter halo,
(NFW plus a clumped component). The parameters were chosen as in 
case 1 above.
The calculation was done for one specific realization of clumps in the 
halo, since we are only interested here in the general features which 
may appear in the power spectrum (the same realization shown in Fig. 1). 

The solid line (crosses) in Fig. 2 is the power spectrum corresponding 
to the smooth profile case (no clumps), while the dashed line (diamonds)  
is the power spectrum for the full dark matter profile (including clumps). 
In both spectra the dipole anisotropy is evident. This anisotropy is due  
to the asymmetric position of the Earth in the galaxy; it arises 
from the fact that the flux from the direction of the galactic center 
far exceeds (by an order of magnitude) that from the antigalactic center. 
As shown in \cite{bm,medina} the present observations are not inconsistent 
with this predicted dipole anisotropy.  
Although not evident in this realization, there are cases in which the 
power spectrum (of the smooth plus clumped models) has some features 
(bumps).  These are generally correlated with the presence 
of a nearby clump of dark matter which gives a substantial contribution
to the flux on a specific angular scale. 

Contrary to the dipole enhancement, which is generic, the amplitude
and frequency of the other bumps completely depends on the specific 
substructure distribution of the Milky Way's halo in which we happen to 
live. It is, therefore, difficult to quantify the likelihood of detecting 
these smaller bumps in the power spectrum of a future experiment. However, 
if such structures in the power spectrum are detected, then they are 
easily understood in the scenario discussed in this letter.

In addition to features in the angular power spectrum (dipole and bumps)
analyses of the composition of events will be precious in constraining 
or confirming the scenario of SHDM in the halo; as discussed above, this  
should be dominated by gamma rays. For this reason, the halo magnetic field 
does not affect our conclusions. (If the proton and gamma ray fractions in 
UHECRs generated by SHDM are comparable \cite{sarkar}, then the magnetic 
field might induce deflections on scales of a degree or so.  
Since we considered slightly larger scales, our conclusions are unlikely 
to change dramatically even in this case.)

The study of the composition, together with an improved 
measure of the spectrum of UHECRs, should nail down the nature of the 
``real'' sources of UHECRs and confirm or rule out the SHDM model.

{\bf Aknowledgments} This work was
supported by the DOE and the NASA grant NAG 5-7092 at Fermilab.

\newpage
{\bf Figure Caption}
\vskip .5cm
{\bf Fig. 1} Map of the ratio of the local flux of UHECRs obtained from a
realistic dark matter distribution in the halo (NFW plus clumps) and the
pure NFW case, for a specific realization of clumps in the halo.

{\bf Fig. 2} Flux power spectrum for the NFW case (solid line with crosses)
and the NFW plus clumped case (dashed line with diamonds). The same realization
of clumps as in Fig. 1 has been used.

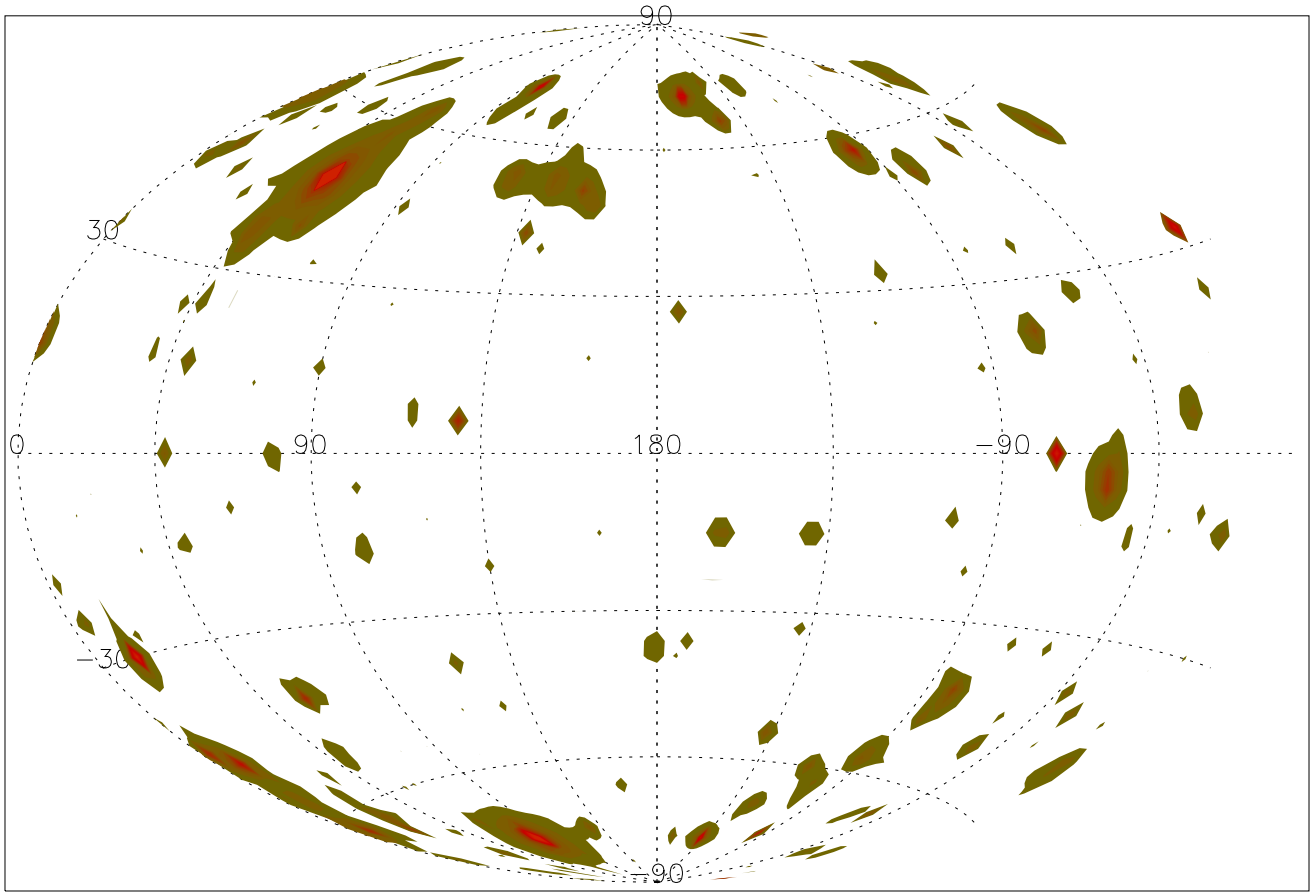
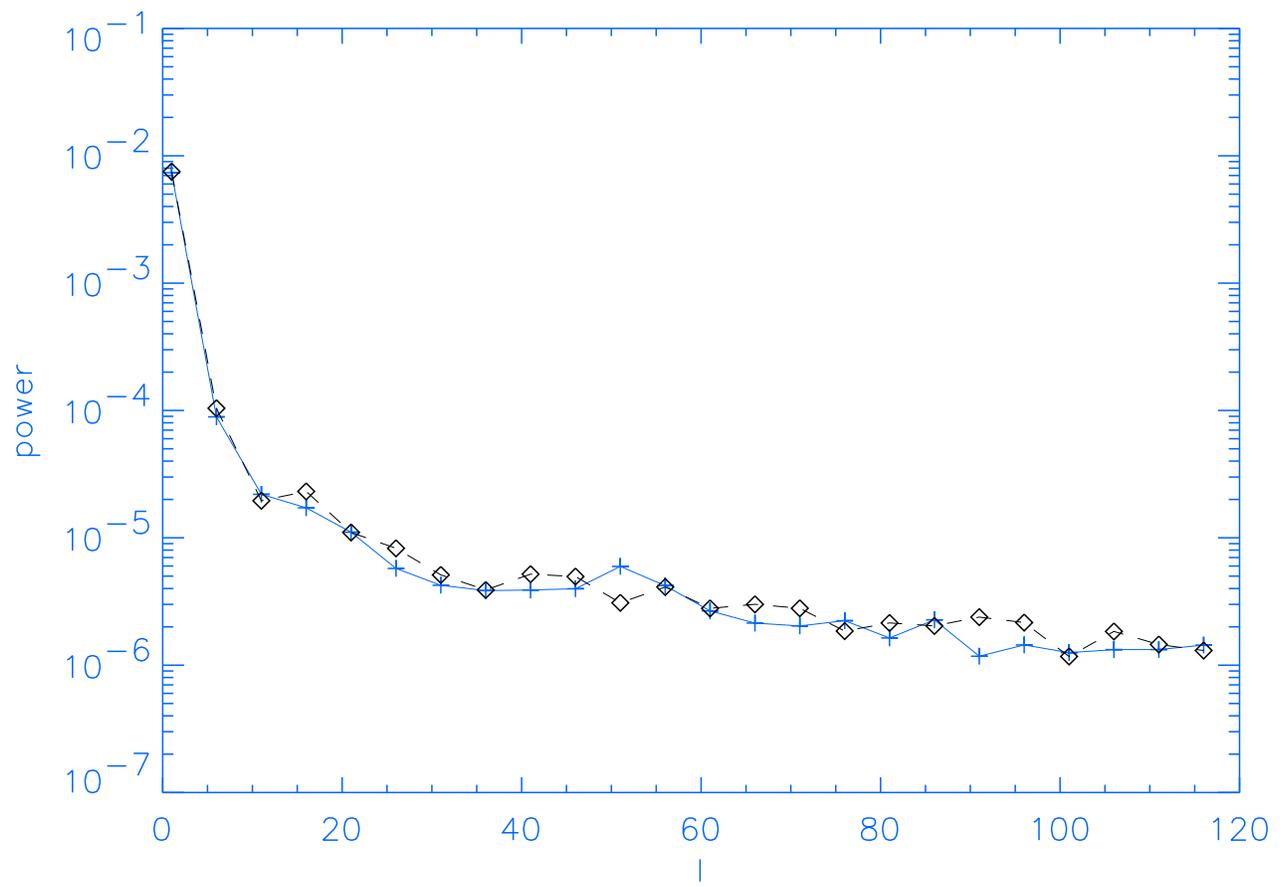
\end{document}